# Detecting Pulmonary Embolism from Computed Tomography Using Convolutional Neural Network

Chia-Hung Yang[a], Yun-Chien Cheng[a, ‡], Chin Kuo[b,c, ‡]

[a] *Department of Mechanical Engineering, College of Engineering, National Yang Ming Chiao Tung University, Hsin-Chu, Taiwan*

[b] *Department of Oncology, National Cheng Kung University Hospital, College of Medicine, National Cheng Kung University, Tainan, Taiwan*

[c] *College of Artificial Intelligence, National Yang Ming Chiao Tung University, Hsin-Chu, Taiwan*

[‡]*The authors contributed equally to this work.*
*Corresponding author: yccheng@nycu.edu.tw , tiffa663@gmail.com

## Abstract

The clinical symptoms of pulmonary embolism (PE) are very diverse and non-specific, which makes it difficult to diagnose. In addition, pulmonary embolism has multiple triggers and is one of the major causes of vascular death. Therefore, if it can be detected and treated quickly, it can significantly reduce the risk of death in hospitalized patients. In the detection process, the cost of computed tomography pulmonary angiography (CTPA) is high, and angiography requires the injection of contrast agents, which increase the risk of damage to the patient. Therefore, this study will use a deep learning approach to detect pulmonary embolism in all patients who take a CT image of the chest using a convolutional neural network. With the proposed pulmonary embolism detection system, we can detect the possibility of pulmonary embolism at the same time as the patient's first CT image, and schedule the CTPA test immediately, saving more than a week of CT image screening time and providing timely diagnosis and treatment to the patient.

# Introduction

Pulmonary embolism (PE) is an obstruction of the pulmonary vasculature caused by the flow of blood clots in the blood vessels, or by the flow of foreign bodies such as tumors, fat masses, or body tissue fragments into the lungs [1]. The clinical symptoms are very diverse and non-specific, which makes it difficult to diagnose. In addition, pulmonary embolism has multiple triggers and is one of the major causes of vascular death. Therefore, if it can be detected and treated quickly, it can significantly reduce the risk of death in hospitalized patients [2-5].

In the detection process of pulmonary embolism, all patients considered by physicians to have suspected pulmonary embolism need to undergo computed tomography pulmonary angiography (CTPA) as the final confirmation. However, the cost of CTPA is high, and angiography requires the injection of contrast agents, which increase the risk of damage to the patient. In addition, potential pulmonary embolism is often undetected in the clinical workup process, and these potential undetected patients have a mortality rate of up to 30%. Therefore, if each patient could be screened for pulmonary embolism immediately and quickly at the time of CT, and CTPA images could be simulated in CT sections with possible pulmonary embolism symptoms, it would not only improve the detection rate of pulmonary embolism, significantly reduce the need for CTPA imaging, reduce the risk of exposure to radiation and contrast, but also effectively shorten the waiting time for patient reports. It can also shorten the waiting time for patient reports.

Therefore, this study will use a deep learning approach to detect pulmonary embolism in all patients who take a CT image of the chest using a convolutional neural network. With the proposed pulmonary embolism detection system, we can detect the possibility of pulmonary embolism at the same time as the patient's first CT image, and schedule the CTPA test immediately, saving more than a week of CT image screening time and providing timely diagnosis and treatment to the patient.

# Method

Chest diseases include PE and many other problems, such as patients with tuberculosis, lung cancer, pneumonia, and even breast cancer. To find out PE patients effectively, we want to build a chest CT wide screening procedure. During chest CT wide screening, we have to also classify patients with other diseases.

Convolutional Neural Network (CNN) is the state of the art in the field of deep learning in classifying medical images, including CT images [6]. We attempted to classify all CT image cross-sections into three categories directly using a convolutional neural network: PE slices, slices with other diseases, and WNL slices. However, because of the great variability in the characteristics of the slices with other diseases, and the serious data imbalance among the categories, which the WNL slices is more than 10 times the amount of data of the PE slices. The lack of focus on PE characteristics and the data imbalance makes the performance of multi class classification using one single model rather unsatisfactory. Moreover, in the clinical application scenario, we would like to accurately exclude WNL slices to reduce the burden of imaging and improve the efficiency of hospital treatment.

Previous research like G Kang et. al. [7] shows that binary classification performs better than ternary classification in the domain of adopting CNNs on CT image classification. To identify possible PE patients from a wide range of screens, we designed a two-stage triage classification model shown in Figure 1. The first stage is a classification network designed to separate with normal limits (WNL) slices from those with diseases. The CT slices that have been classified as having a disease will be sent to the second stage for further classification. The second stage is another classification network that identifies slices that may have PE. The two-stage model performs better on filtering out PE slices from the whole dataset and provides other potential lung diseases from the output of the first stage model for physicians' inference.

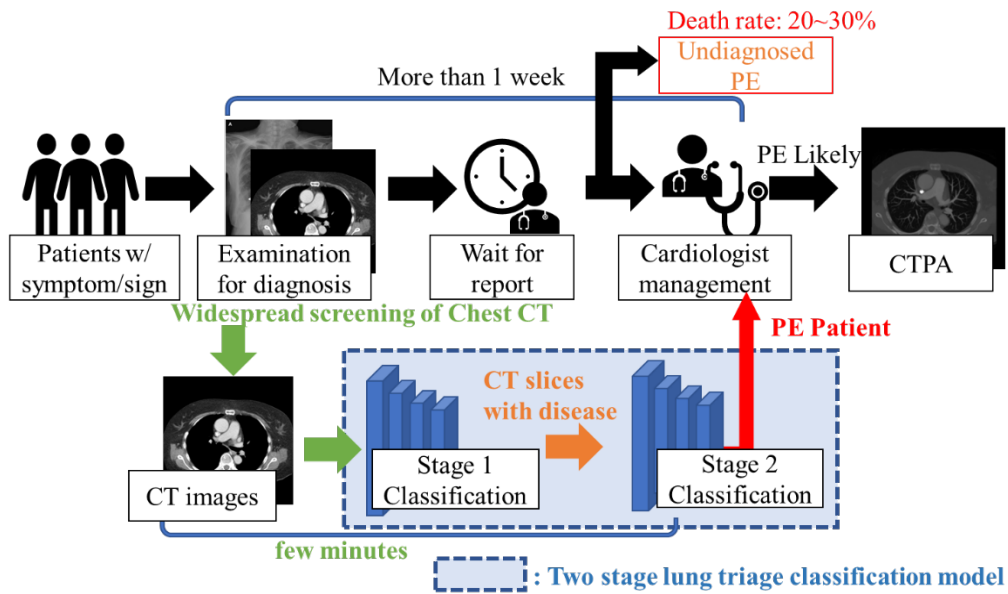

Fig 1. PE diagnose process and clinical usage of the two-stage lung triage classification. The black line is the original process of diagnosing PE patients which is inefficient and cannot perform a widespread screening. The two-stage lung triage classification model not only shorten the PE diagnose process time but also avoid the patients from not being diagnosed.

**Dilated Residual Network with Block Attention Model**

The characteristics of chest CT images are usually quite complex, and the structural differences between each patient are also quite large. Take PE as an example. PE can occur in any area with blood vessels throughout the lungs. PE may occur in the central large blood vessels of the peripheral capillaries, and the size of the resulting area is also quite different. Therefore, a convolutional neural network with a fixed convolution kernel is not easy to achieve a satisfactory classification effect in a lightweight architecture, but a convolutional neural network that is too deep is not suitable for clinically use. To achieve a model that is both accurate and clinically usable, we chose dilated residual network (DRN) [8] as the backbone of the convolutional neural network and supplemented it with attention mechanism to enhance the model's ability to focus on features.

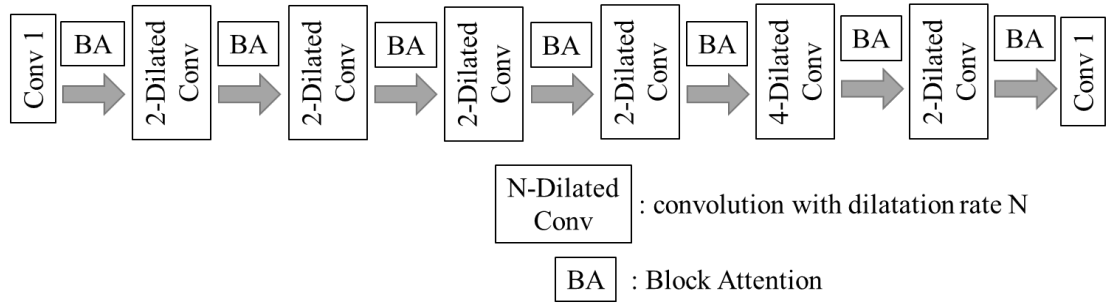

Fig 2. Dilated Residual Network with Block Attention Model(CBAMDRN). The Classification model backbone used in the study, the DRN backbone was based on DRN-d-50 and with block attention attached between each dilated convolution blocks.

DRN was proposed by Fisher Yu et. al. [8] to overcome the loss of spatial acuity caused by resolution reduction. Convolution neural networks reduce the resolution of an image until the image is represented by a feature map, and make the spatial structure detailed and discernible. Dilated convolution is a type of convolution that inflates the kernel by inserting holes between the kernel elements. With these holes inserted in the kernel, the spatial structure will not vanish while the convolution kernel reduced the resolution. As a result, DRN better preserves spatial resolution in convolutional networks for image classification.

As the dilated convolution backbone preserved the spatial resolution, the model should find out the region of interest of lung diseases. Xiao Chen et. al. [9] proposed a cascade attention module to capture the non-local feature dependencies and help the deep neural network focus on the lesion area. A state-of-the-art attention mechanism is block attention. Sanghyun Woo et. al. [10] attach channel-wise attention with spatial-wise attention as block attention and integrated with a ResBlock in ResNet. We inserted the block attention between every layer of the DRN Blocks and created a CBAMDRN model as Figure 2. to focus on lesion areas of lung diseases.

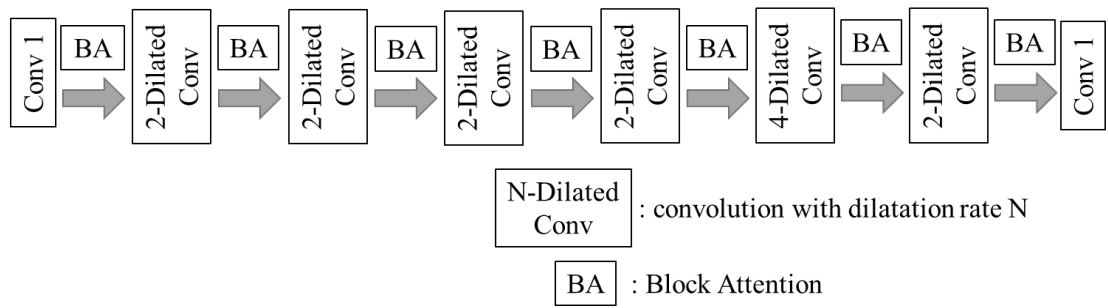

Fig 2. Dilated Residual Network with Block Attention Model(CBAMDRN). The Classification model backbone used in the study, the DRN backbone was based on DRN-d-50 and with block attention attached between each dilated convolution blocks.

**Multi-window level input (MWL) structure**

To retain the maximum information, our classification model takes the original Dicom image as input. The original CT volume range of Dicom images is very large, approximately between -1000 and 3000. The original CT volume information is directly related to how tissue interacts with X-rays. Doctors need suitable filters to read CT images, and different filters are needed to observe different diseases. PE lesions are the priority target that our model should identify, so we designed a filter that can best filter out the pulmonary blood vessels to better observe the PE characteristics. The image model observed on this vessel filter is named the vessel window input. Some diseases, such as pneumonia, exist in lung tissue. Therefore, we have to use two other filters commonly used by doctors to observe lung diseases, namely the lung window input and the mediastinal window input.

We purposed a Multi-window level input model to get features from three different input windows respectively. The structure of the multi-window level input model is shown in Figure 3. Three different input levels will go through different CBAMDRN models to extract different feature maps and then be concated by a fully connected layer as a single prediction output. We found out the mediastinal window input has less impact on our model to diagnose PE and lung diseases, so we created a lung window level and vascular window level only, double window level input model (DWL) to lower the interrupting noise from the mediastinal window input.

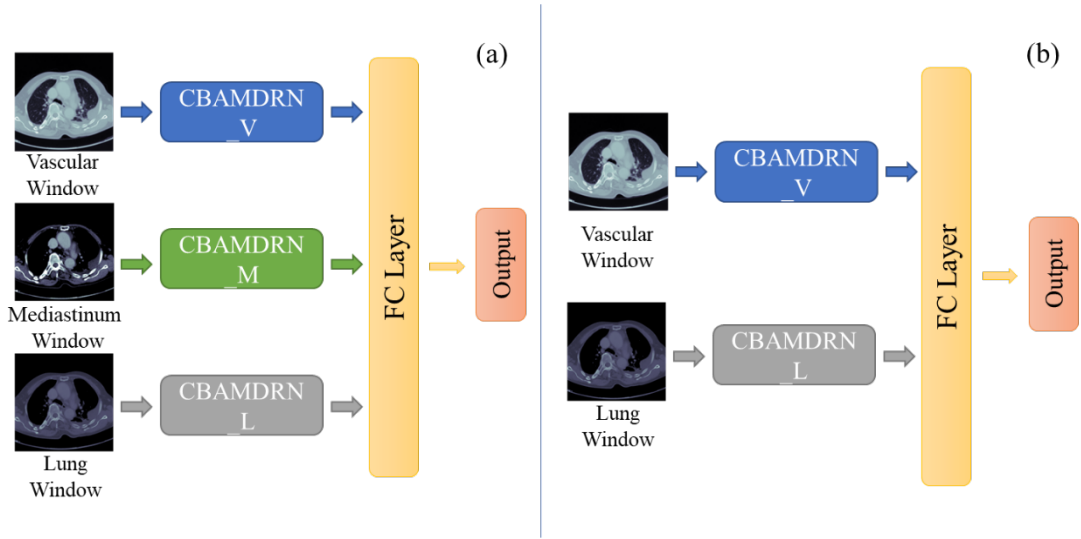

Fig 3. Multi-window level input model. Triple window level input (TWL) model (a) includes three different window level input. Double window level input (DWL) model (b) only includes Vascular window level and lung window level. CBAMDRN_V is a CBAMDRN model having a vascular window level input, while CBAMDRN_M and CBAMDRN_L takes mediastinum window level and lung window level respectively.

**Data Preprocess**

In order to preprocess the data of the CT input model, pixel data were extracted from the original Digital Imaging and Communication in Medicine (DICOM) format for each CT examination. The image was cropped to a size of 400*400, focusing only on the lung area. Three different Hounsfield Unit (HU) ranges were chosen to look at different aspects of the lung scan is shown in Table 1. The first window level is the lung window, the HU is cropped to the range of -400 to 1500, the second window level is the mediastinal window, the HU is cropped to the range of 40 to 400, and the last window level is a wider window level for viewing lungs and blood vessels, which was cropped from -1000 to 1600. All three windows are normalized to zero center.

Table 1. CT image with different window view. The default window view is designed to observe vessels while lung tissue is still visible. Mediastinum window level and lung window level are often an inference for clinical diagnoses.

|  | Vascular window | Mediastinum window | Lung window |
|---|---|---|---|
| Image | 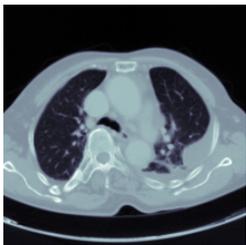 | 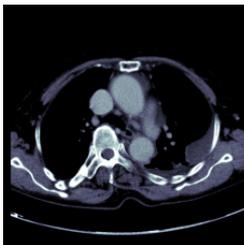 | 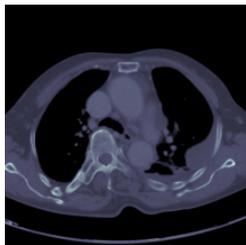 |
| HU value range | 0 - 650 | 40 - 400 | -400 - 1500 |

After filtering the HU window, another crop is performed to focus at the lung area. A size of 300*180 in the central is chosen as a fixed cropping of the lung but might lost a lot of information due to the variety between the patients, the size of lung and the lung location are often different. We proposed the dynamic lung cropping method to overcome this problem. To better define the lung area, the lung CT pixel data were binarized and made the lung area two huge black space in the middle, and then fill back the original pixel data after cropping out the black area by a bounding box. This dynamic lung cropping method successfully cropped out the lung area and not causing a heavy computational burden.

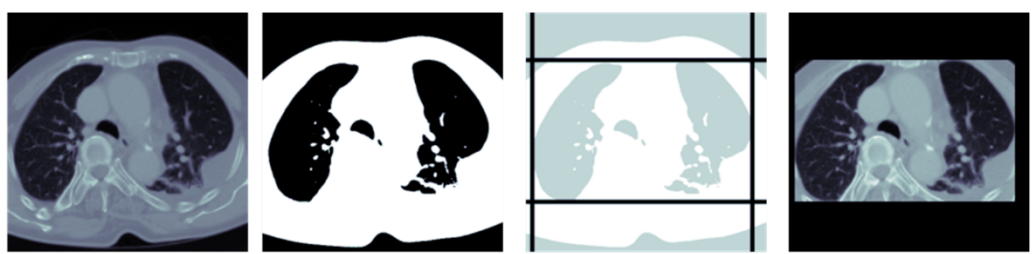

Step 1. Center crop 400*400  
Step 2. Binarization  
Step 3. Find lung boarder  
Step 4. Crop the lung

Fig 4. The dynamic lung cropping process. To focus on lung diseases, cropping out the lung from the whole CT slice can avoid including unnecessary interference factors. The dynamic lung cropping process cropped the lung area precisely and does not cause heavy computation burden.

## Experiment Results and Discussion

**Dataset**

The Chest CT images used in this study were collected from National Cheng Kung University Hospital. (IRB No: B-ER-108-380) The dataset includes 53 different patients' lung CT images. The dataset was divided into three different classes, PE, other lung diseases, and WNL. And the number of patients with PE, WNL, and those with other diseases is 13, 14, 26, respectively. Most of the patients with diseases don't have lesions throughout the whole lung, which means we will have to pick out the slices that contain PE or other lung disease lesions. We only select CT image slices containing disease lesions to be classified as PE or other diseases, and the remaining slices are classified as WNL slices. There are a total 2704 slices in the dataset, and we have 167, 655, 1882 slices for PE, other diseases, and WNL respectively. The dataset was divided into the training set, validation set, and testing set by 5:3:2. The raw chest CT Dicom files directly input our model to do the classification task.

The WNL images used in this experiment are specially selected images of breast cancer patients. In fact, in the medical imaging application scenario, there are almost no WNL patients in the chest CT images collected by hospitals. To make the model recognize normal lungs, a larger number of WNL patients are added to improve the accuracy of judgment.

**First stage classification**

In the first stage of our classification model, our goal is to find out the CT slices with diseases. The input image were all filtered by vascular window level, and we compared different convolutional neural networks to find out the one best suit our dataset. The models we chose are ResNet 50 [11], DRN, CBAMDRN and Efficient Net b5 [12]. And we also compare the results under a fixed cropping and the dynamic cropping method we purposed. We evaluate the results with accuracy and sensitivity.

$$\text{Accuracy} = \frac{Total\ number\ of\ correctly\ classified\ CT\ slices}{Total\ number\ of\ CT\ slices}$$

$$\text{Disease sensitivity} = \frac{Total\ number\ of\ correctly\ calssified\ Disease\ slices}{Total\ number\ of\ Disease\ slices}$$

$$\text{WNL sensitivity} = \frac{Total\ number\ of\ correctly\ calssified\ WNL\ slices}{Total\ number\ of\ WNL\ slices}$$

As we can see in Table 2, the dynamic cropping method outperformed the fixed cropping method in every model, and the CBAMDRN performs the best on overall accuracy and WNL sensitivity. Although Efficient Net performed closely to the CBAMDRN model, the parameter of Efficient Net is 26 times larger than the CBAMDRN model. The huge parameter size makes Efficient Net inefficient in clinical use.

Table 2. The class-wise per slice accuracy and sensitivity of fix crop and dynamic lung cropping in the Stage 1 classification.

**WNL/Disease**

| Model | Cropping method | Accuracy | WNL sensitivity | Disease sensitivity |
|---|---|---|---|---|
| ResNet50 [13] | Fix crop | 93.5% (507/542) | 96.6% (364/377) | 86.7% (143/165) |
| | Dynamic crop | 93.7% (508/542) | 94.4% (356/377) | 92.1% (152/165) |
| DRN [15] | Fix crop | 93.0% (504/542) | 96.8% (365/377) | 84.2% (139/165) |
| | Dynamic crop | 96.0% (520/542) | 96.2% (363/377) | 95.2% (157/165) |
| Efficient Net [16] | Fix crop | 93.4% (506/542) | 96.8% (365/377) | 85.5% (141/165) |
| | Dynamic crop | 96.0% (520/542) | 95.2% (359/377) | 97.5% (161/165) |
| CBAMDRN (Ours) | Dynamic crop | 96.1% (521/542) | 97.1% (366/377) | 93.9% (155/165) |

Next, we tested the performance of the multiple window level input model in Table 3. The result shows that the Double window level (DWL) input model with vascular window and lung window performs the best in our first stage classification model. These three models share the same WNL sensitivity but DWL outperforms the others in disease sensitivity.

Table 3. The class-wise per slice accuracy and sensitivity of Multi-window level input model with dynamic lung cropping.

| Input | Accuracy | WNL sensitivity | Disease sensitivity |
|---|---|---|---|
| **Vascular window** | **96.1%** (521/542) | **97.1%** (366/377) | **93.9%** (155/165) |
| **DWL** | **96.8%** (525/542) | **97.1%** (366/377) | **96.3%** (159/165) |
| **MWL** | **95.6%** (518/542) | **97.1%** (366/377) | **92.1%** (152/165) |

For the qualitative analysis, we apply the gradient-weighted class activation mapping (Grad-CAM) [13] to explain where the model focus on. We performed the grad cam on MWL to find out why adding mediastinum window level drops the performance on the case of PE classification. We found out that the model can't focus on any important features in mediastinum window level, and the grad cam shows a messy gradient while having a significant focus on vascular and lung level. And the gradients on vascular and lung window level both includes PE lesion.

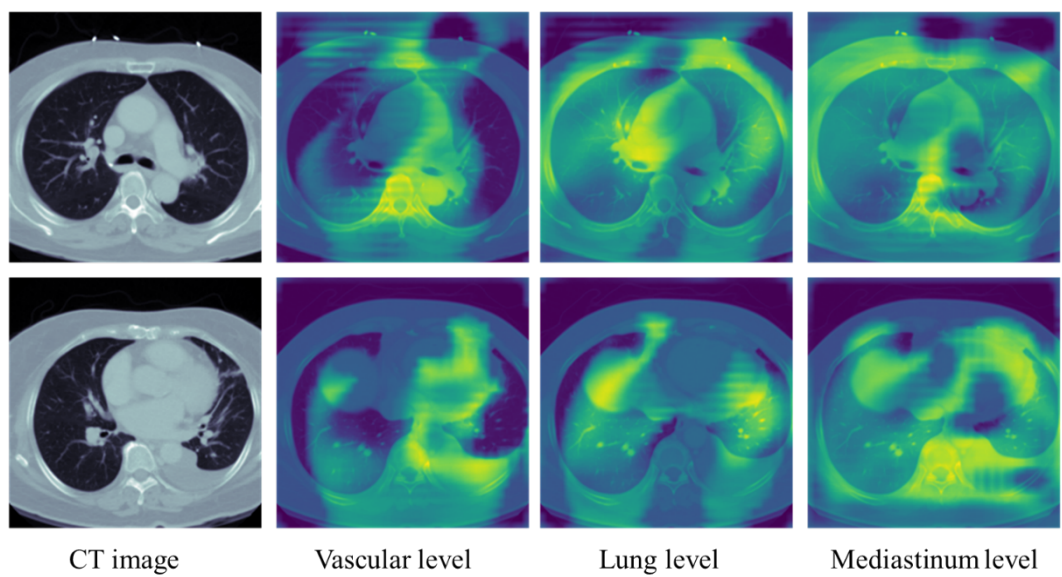

Fig 5. The Grad-CAM of each Multi-window level input model.

Next, we tested the results per patient to simulate the clinical use of these models.

If a patient has one single slice classified as disease, the patient will be classified as disease. In Table 4, we can find out that Efficient Net, CBAMDRN, and DWL models all perform perfectly in classifying disease patients from WNL patients.

Table 4. The class-wise per patient accuracy and sensitivity in the Stage 1 classification.

**WNL/Disease**

|  | Cropping method | Accuracy | Disease sensitivity | WNL sensitivity | *WNL per slice sensitivity |
|---|---|---|---|---|---|
| **ResNet50** | Fix crop | 50% (4/8) | 100% (4/4) | 0% (0/4) | 81.6% (200/245) |
| **DRN** | Fix crop | 50% (4/8) | 100% (4/4) | 0% (0/4) | 50.6% (124/245) |
| **Efficient Net** | Dynamic crop | 100% (8/8) | 100% (4/4) | 100% (4/4) | 100% (245/245) |
| **CBAMDRN (Ours)** | Dynamic crop | 100% (8/8) | 100% (4/4) | 100% (4/4) | 100% (245/245) |
| **DWL DRN (Ours)** | Dynamic crop | 100% (8/8) | 100% (4/4) | 100% (4/4) | 100% (245/245) |

**PE classification**

The second stage of the classification model is to classify PE slices from slices with other diseases. The input images were all filtered by vascular window level in single window lever input models. In Table 5, we compared different convolutional neural networks with DWL and MWL together to find out the one that recognizes PE feature best. And we also compare the results under a fixed cropping and the dynamic cropping method we purposed in this stage too. We evaluate the results with accuracy and sensitivity, and the better the PE sensitivity is the greater chance the possible PE patient has been discovered and cured. The result shows that although ResNet 50 performed perfectly on no-PE sensitivity, it seems like the model can't classify PE slices correctly. The DWL model had a 99.6% overall accuracy and performed perfectly on classifying PE slices.

$$\text{PE sensitivity} = \frac{Total\ number\ of\ correctly\ calssified\ PE\ slices}{Total\ number\ of\ PE\ slices}$$

$$\text{NoPE sensitivity} = \frac{Total\ number\ of\ correctly\ calssified\ other\ disease\ slices}{Total\ number\ of\ other\ disease\ slices}$$

Table 5. The class-wise per slice accuracy and sensitivity of the second stage classification.

WNL/Disease

|  | Cropping method | Accuracy | Disease sensitivity | WNL sensitivity | *WNL per slice sensitivity |
|---|---|---|---|---|---|
| ResNet50 | Fix crop | 50% (4/8) | 100% (4/4) | 0% (0/4) | 81.6% (200/245) |
| DRN | Fix crop | 50% (4/8) | 100% (4/4) | 0% (0/4) | 50.6% (124/245) |
| Efficient Net | Dynamic crop | 100% (8/8) | 100% (4/4) | 100% (4/4) | 100% (245/245) |
| CBAMDRN **(Ours)** | Dynamic crop | 100% (8/8) | 100% (4/4) | 100% (4/4) | 100% (245/245) |
| DWL DRN **(Ours)** | Dynamic crop | 100% (8/8) | 100% (4/4) | 100% (4/4) | 100% (245/245) |

Finally, we tested the results again but per patient to simulate the clinical use of these models. If a patient has one single slice classified as PE, the patient will be classified as PE. In Table 6, we can find out that CBAMDRN performs likely as DWV model, but DWV model only misclassify one single image in the whole testing set.

Table 6. The class-wise per patient accuracy and sensitivity of the Stage 2 classification.

WNL/Disease

|  | Cropping method | Accuracy | Disease sensitivity | WNL sensitivity | *WNL per slice sensitivity |
|---|---|---|---|---|---|
| ResNet50 | Fix crop | 50% (4/8) | 100% (4/4) | 0% (0/4) | 81.6% (200/245) |
| DRN | Fix crop | 50% (4/8) | 100% (4/4) | 0% (0/4) | 50.6% (124/245) |
| Efficient Net | Dynamic crop | 100% (8/8) | 100% (4/4) | 100% (4/4) | 100% (245/245) |
| CBAMDRN **(Ours)** | Dynamic crop | 100% (8/8) | 100% (4/4) | 100% (4/4) | 100% (245/245) |
| DWL DRN **(Ours)** | Dynamic crop | 100% (8/8) | 100% (4/4) | 100% (4/4) | 100% (245/245) |

We took a look at the only misclassified image shown as Figure 6 and look into the grad cam. We found out that the image is highly likely to have PE by only observing this single slice, but after double checking the whole patient's CT slices, this slice isn't PE. The circled area is the area likely to be diagnosed as PE, and it greatly fits the Grad-CAM on both window level. This shows that the model is focusing on the same feature physician observing.

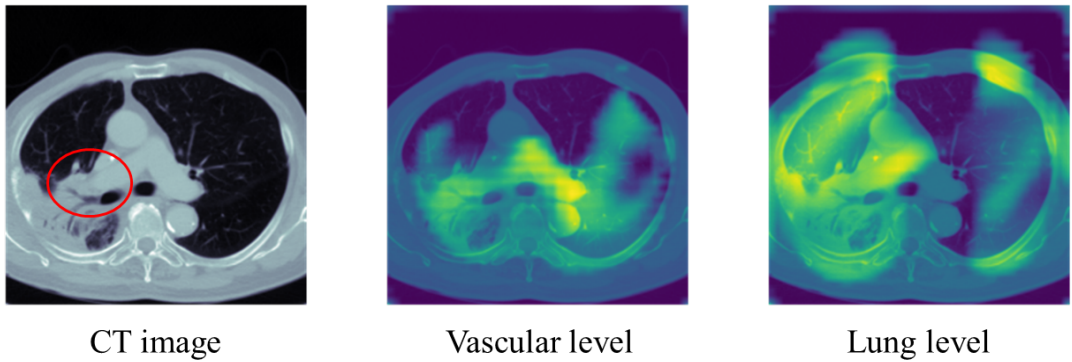

      CT image        Vascular level        Lung level

Fig 6. The CT slice misclassified as PE and the Grad-CAM of the input windows.

## Conclusion

In our current study, the two-stage classification network simplifies the multiclass classification problem into two dichotomous classification problems, which reduces the complexity of the problem while allowing the deep learning model to focus on different features in two different stages to achieve a high accuracy of classification. In addition, it is found that by using our designed multi-window input images with null convolution and attention mechanism, we can achieve 96.8% and 99.4% accuracy in the recognition of lung computed tomography images containing diseases or in the selection of more urgent pulmonary embolism patients from many cross-sections containing different lung-related diseases, respectively. We also achieved a 100% PE detection rate in patient-based clinical classification. However, the images we trained were all from Cheng Kung University Hospital. For further clinical testing, we need to increase the number of datasets from different hospitals to verify whether our classification network can maintain the same classification accuracy in different datasets.

## Reference:
1. Goggs, Robert, et al., 2009, "Pulmonary thromboembolism." Journal of Veterinary Emergency and Critical Care 19.1: 30-52.
2. Stein, Paul D., et al., 2009, "Diagnosis of pulmonary embolism in the coronary care unit." The American journal of cardiology 103.6: 881-886.
3. Apfaltrer, Paul, et al., 2011, "CT Imaging of Pulmonary Embolism: Current Status." Current Cardiovascular Imaging Reports 4.6: 476-484.
4. Page, Patient, et al., 2006, "Effectiveness of managing suspected pulmonary embolism using an algorithm combining clinical probability, D-dimer testing, and computed tomography." Jama 295.2 (2006): 172-179.
5. Migneault, David, et al., 2015, "An unusual presentation of a massive pulmonary embolism with misleading investigation results treated with tenecteplase." Case reports in emergency medicine 2015.
6. Albawi, Saad, et al, 2017, "Understanding of a convolutional neural network." 2017 International Conference on Engineering and Technology (ICET).
7. Guixia Kang, et al, 2017, "3D multi-view convolutional neural networks for lung nodule classification." PlosOne.
8. Fisher Yu, et al, 2017, "Proceedings of the IEEE Conference on Computer Vision and Pattern Recognition (CVPR)," 2017, pp. 472-480
9. Xiao Chen et al, 2019, "A Cascade Attention Network for Liver Lesion Classification in Weakly-Labeled Multi-phase CT Images." MICCAI Workshop on Domain Adaptation and Representation Transfer.
10. Sanghyun Woo et al, 2018, "CBAM: Convolutional Block Attention Module." Proceedings of the European Conference on Computer Vision (ECCV), pp. 3-19.

11. A. Sai Bharadwaj Reddy et al., 2019 "Transfer Learning with ResNet-50 for Malaria Cell-Image Classification" International Conference on Communications and Signal Processing.

12. M Tan et al., 2019 "EfficientNet: Rethinking Model Scaling for Convolutional Neural Networks," arXiv:1905.11946.

13. Ramprasaath R et al., 2017, "Grad-CAM: Visual Explanations From Deep Networks via Gradient-Based Localization" Proceedings of the IEEE International Conference on Computer Vision (ICCV), pp. 618-626